\documentclass[aps,twocolumn,amsmath,amssymb,
prl,showpacs,showkeys,array]{revtex4}
\usepackage[dvips]{graphicx}
\usepackage{psfrag}

\begin{document}
\title{Discrete breathers in Fermi-Pasta-Ulam lattices }
\author{S. Flach and A. Gorbach}
\affiliation{Max-Planck-Institut f\"ur Physik komplexer Systeme, N\"othnitzer
Strasse 38, D-01187 Dresden, Germany}
\date{\today}

\begin{abstract}
We study the properties of spatially localized and time-periodic excitations - discrete breathers -
in Fermi-Pasta-Ulam (FPU) chains. We provide a detailed analysis of their spatial
profiles and stability properties.
Especially we demonstrate that the Page mode is linearly stable for symmetric FPU potentials.
A resonant interaction between a localized and delocalized perturbations causes
weak but finite strength instabilities for asymmetric FPU potentials. This interaction
induces Fano resonances for plane waves scattered by the breather. Finally we analyse
the interplay between energy thresholds for breathers in the presence of strongly asymmetric FPU potentials
and the corresponding profiles of the low-frequency limit of breather families. 
\end{abstract}


\maketitle

{\bf The Fermi-Pasta-Ulam (FPU) paradox was observed fifty years ago. The surprising
finding was a localization of energy in the reciprocal $q$-space of 
a model with discrete translational invariance, despite the presence of
interaction between extended normal modes. Thirty three years later Sievers, Takeno and Kisoda
reported on the observation of energy localization in real space for the same class of FPU models,
which is as surprising since these excitations, called discrete breathers or intrinsic localized modes,
 violate the underlying discrete translational symmetry of the model.
The past decade has witnessed a tremendous progress in the theory and applications of
discrete breathers, which goes much beyond the scope of the original FPU frame. 
We use the modern theory of discrete breathers to investigate the properties of these
solutions in FPU models, paying special attention to the issues of stability, resonances,
wave scattering and energy thresholds.  }

\section{Introduction}

The celebrated Fermi-Pasta-Ulam (FPU) model was introduced fifty
years ago in order to study the process of equilibration
of energy among normal modes due to mode-mode interactions \cite{fpu55}.
It can be viewed as a toy version of a model
describing the dynamics of lattice vibrations of perfect crystals.
The original FPU model reduced space dimension to one,
and attributed one degree of freedom to each lattice site.
It is worth to mention that in the absence of mode-mode interactions,
that model is often used in solid state physics textbooks to
explain the basic features of
phonons and is then called
a {\it monoatomic chain} \cite{ck96}. Some of these textbooks deal also with
mode-mode interactions, coining them {\it anharmonic corrections}
(since they appear as anharmonic terms in the Taylor expansion of
the potential energy of the lattice with respect to atomic displacements).
Notably these anharmonic terms are used to explain thermal expansion
of crystals, among many other features.
It is also worth to stress that solid state physicists did not
seriously question the assumed fact that in thermal equilibrium
the energy {\it is} equipartitioned among the normal modes -
due to the bulk of accumulated experimental evidence for that fact.
The more time-resolved spectroscopical methods
advance, the more questions arise concerning these topics when studying
lattice dynamics in the presence of strong nonlinearities (i.e. mode-mode
interactions).
Returning to the FPU model calculations, they revealed the surprising
fact that at least for some cases (for some sets of initial conditions)
the evolution of the FPU model showed {\it no} equipartioning
among the normal modes \cite{fpu55}. In other words, the FPU {\it paradox} seems
to consist primarily of the observation of {\it localization} of 
energy in a few normal modes, despite the presence of interaction
between all modes, which would be capable of distributing the energy
among all normal modes of the system \cite{ic66}.

%
\begin{figure}
\vspace{20pt} \centering
\includegraphics[width=0.9\columnwidth]{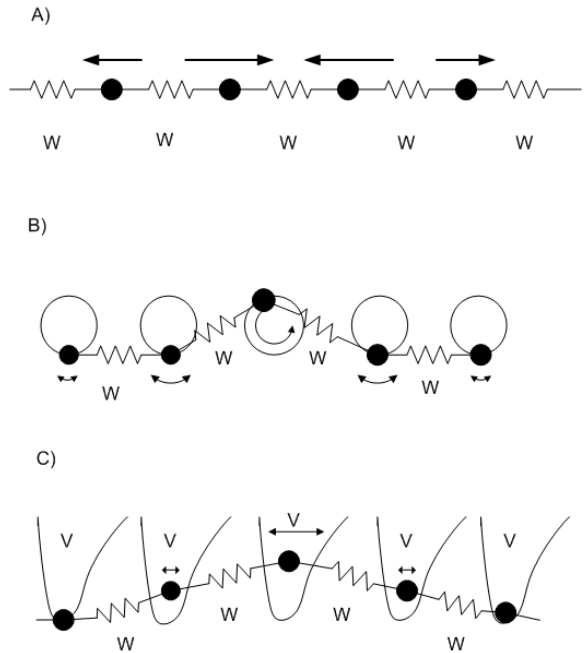}\\
\caption{A schematic representation of different types of discrete breathers:
(a) acoustic FPU breather; (b) acoustic rotobreather; (c) optical breather
(for details see e.g. \cite{DB-REVIEWS}).}
\label{figbreather_types}
\end{figure}

While the modern view on this FPU paradox observation is for sure
discussed in other contributions to this focus issue, we will
not dwell on that further. In what follows, we will discuss another
interesting aspect of the dynamics in the FPU model and its
various generalizations. This concerns localization in {\it real space}.
In other words, we will show that FPU models allow for solutions of
the corresponding equations where the energy is not equally shared among
the {\it local} constituents (using the solid state physics language,
among the different atoms).
These solutions are coined {\it Intrinsic Localized Modes} (ILM), or
{\it Discrete Breathers} (DB). Whatever the nomenclature,
ILMs/DBs turn out to be generic
to a much larger class of {\it Hamiltonian lattices}, and FPU models
together with their generalizations represent one of the subclasses
of these lattices (see Fig.\ref{figbreather_types} for a schematic
representation of various DBs).
ILMs/DBs are time-periodic and spatially localized
solutions, and exist thanks to the interplay between nonlinearity
and discreteness \cite{DB-REVIEWS}. Many studies of DBs
have been successfully launched,
on such topics as
rigorous existence proofs, dynamical and structural
stability and computational methods of obtaining DBs
in classical models as well as their quantum aspects.
In addition DBs have been detected and studied experimentally
in such different systems as interacting Josephson junction systems
\cite{binder00a},
coupled nonlinear optical waveguides \cite{eisenberg98},
lattice vibrations in
crystals \cite{swanson99}, antiferromagnetic structures \cite{schwarz99},
micromechanical cantilever arrays \cite{sato03}, Bose-Einstein
condensates loaded on optical lattices \cite{BEC}, layered high-$T_c$
superconductors \cite{hightc}. DBs are predicted also
to exist in the dynamics of dusty plasma crystals \cite{plasma}.
It is an equally interesting question of why ILMs/DBs have not been
properly discussed when studying the lattice dynamics of crystals
until very recently.
We leave the answer to this question to the experienced and educated
experts of that field.

\section{Setting the stage and early results}

We will consider the following
class of one-dimensional Hamiltonian chains 
\begin{equation}
H=\sum_l \left[ \frac{1}{2} p^2_l + W(x_l - x_{l-1}) \right]
\;,
\label{1-1}
\end{equation}
where $x_l$ describes the scalar displacement of a particle (atom)
from its equilibrium position, $p_l = \dot{x}_l\equiv dx_l/dt$ is its
conjugated momentum (velocity),
$l$ denotes the number of the particle,
and $W(x)$ is the interaction
potential between nearest neighbours. The Hamiltonian $H$ is assumed
to take only nonnegative values for small values
of the displacements and velocities, i.e. the potential $W$ and its
first derivative $W'$ vanish for zero displacements $W(0)=W'(0)=0$
and the second derivative $W''$ is positive for small displacements
$W''(0) > 0$.  
The Hamiltonian equations of motions $\dot{x}_l = \partial H / 
\partial p_l\;,\;\dot{p}_l = -\partial H / \partial x_l$
lead to the following
set of coupled differential equations:
\begin{equation}
\ddot{x}_l = -W^{\prime}(x_l - x_{l-1}) + W^{\prime}(x_{l+1}-x_l)\;.
\label{1-2}
\end{equation}
Let us expand the function $W$ in the following series
\begin{equation}
W(x) = \sum_{m=2}^{\infty} \frac{\phi_m}{m} x^m\;.
\label{1-3}
\end{equation}
If the potential $W(x)$ is symmetric $W(x)=W(-x)$ it follows
$\phi_{2m+1}=0$ for all positive integers $m$.

The celebrated FPU models \cite{fpu55} are obtained by choosing nonzero
values for $\phi_{2,3,4}$ and zeroing all the others $\phi_{m > 4}=0$.
Thus we may consider the FPU models as low amplitude expansions
of the more general model class (\ref{1-1}).
FPU models take into account the two first anharmonic corrections $\phi_{3,4}$
to the harmonic term $\phi_2$. The case of a symmetric potential
for FPU models is thus obtained by assuming $\phi_3=0$.
Note that anharmonic terms in the Hamiltonian correspond to nonlinear
terms in the equations of motion here. The equations (\ref{1-2}) conserve
both the total energy $H$ (\ref{1-1}) as well as the {\it total mechanical
momentum} $P = \sum_l p_l$. Without loss of generality we will consider
below the case $P=0$ only.

\begin{figure}[htb]
\vspace{20pt} \centering
\includegraphics[angle=270, width=0.95\columnwidth]{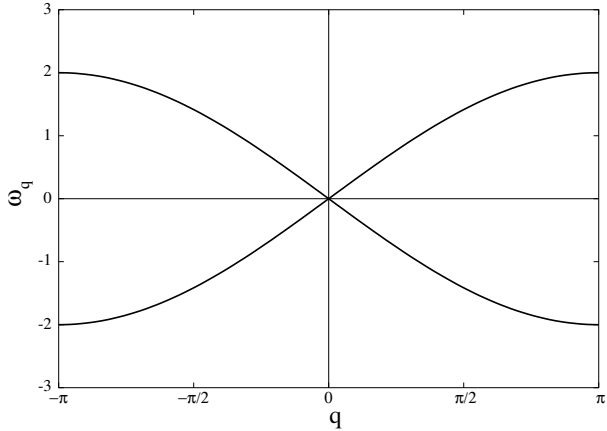}\\
\caption{Dependence of $\omega_q$ on $q$ for $\phi_2=1$.
Vertical and horizontal lines mark the values $q=0$ and $\omega_q=0$.}
\label{fig1}
\end{figure}

If we restrict our consideration to very small amplitudes and velocities
only, we may neglect all nonlinear terms from the equations of motion,
assuming $\phi_{m > 2}=0$. The solution of the corresponding
{\it linear} coupled differential equations
\begin{equation}
\ddot{x}_l = \phi_2(x_{l+1} + x_{l-1} - 2x_l)
\label{1-4}
\end{equation}
can be written in the form of a superposition of
plane waves each given by
\begin{equation}
x_l(t) = A_q \cos (\omega_q t - ql + B_q)
\label{1-5}
\end{equation}
where $q$ is the wave number, $\omega_q$ the plane wave frequency,
and $A_q$ and $B_q$ are integration constants.
The dispersion relation (Fig.\ref{fig1})
\begin{equation}
\omega_q = \pm 2\sqrt{\phi_2} \sin\left(\frac{q}{2}\right)
\label{1-6}
\end{equation}
is periodic in $q$
and is characterized by an upper bound
$|\omega_q| \leq \omega_{\pi}
\equiv 2\sqrt{\phi_2}$.
%
Moreover for small values of $q$
$|\omega_q| \approx \sqrt{\phi_2}q$. Such an {\it acoustic}
type of dispersion at small values of $q$ is intimately connected
to the above mentioned conservation of the total mechanical momentum $P$.
Any spatially localized initial
excitation on the lattice will ultimately disperse due to
the $q$-dependence of the {\it group velocity} 
\begin{equation}
v_q = \frac{d \omega_q}{d q} = \pm \sqrt{\phi_2}\cos \frac{q}{2}
\;.
\label{1-7}
\end{equation}
Taking into account weak anharmonic/nonlinear effects usually
leads to a renormalization of plane wave properties such as
frequencies and lifetimes. However no principal
change in the behaviour of a spatially localized initial condition is expected
from that perspective. 

Some rigorous results exist for
the study of stability of certain plane waves. Indeed, the corresponding
modulational instability (cf. \cite{bb83},\cite{kp94},\cite{sp94},\cite{sf96},\cite{pr97} and 
\cite{dkpr05})
is responsible for the appearance of localized structures.
However
the linearization of the phase space flow around the plane wave 
in these studies 
limits possible conclusions about the appearance of strongly
localized ILMs or DBs.

A nonperturbative and qualitatively new
property of the evolution of spatially localized excitations
takes place if the nonlinear terms become essential.
Instead of observing a decay in space, robust and seemingly exact
spatially localized vibrational excitations have been observed.
Early results on ILMs or DBs in FPU chains have been obtained by
Sievers, Takeno and Kisoda \cite{st88}.
These studies as well as their followups \cite{followups}
provide three types of observations and results, among others.

First, they report on numerical
evidence of the existence of long lived localized excitations,
with lifetimes much larger than the typical oscillation times
$\sim 2\pi/\omega_{\pi}$. 
The internal frequencies of such ILMs are {\it outside} the spectrum
$\omega_q$. In Figs.\ref{fig2},\ref{fig3},\ref{fig4} 
some of these excitations are shown
(and will be discussed in more detail below).
In fact, as will be discussed in the next sections, these excitations
are {\it exact} solutions of the equations of motion.

Second, approximate numerical
solutions were obtained using the {\it Rotating Wave Approximation}
which implies that a time-periodic solution is constructed taking
into account only its first harmonics frequency contribution
(and optionally also a dc component) and
neglecting higher harmonics. The recipe is to use the ansatz
$ x_l(t)=c_l + a_l \cos (\Omega_b t)$, to insert it into
the equations of motion (\ref{1-2}) and to neglect all terms
with higher harmonics which appear due to the nonlinear terms.
The resulting set of coupled algebraic equations for the coefficients
$c_l,a_l$ can be solved numerically for finite numbers of sites,
and yields solutions similar to those observed in numerical simulations.
This approximation is strictly speaking valid only for small
amplitudes of the displacements. In reality however it may serve
as a good estimate to exact solutions for rather large amplitudes
as well.

And third, for quite long simulation
times {\it moving} ILMs have been observed, which in addition
of being characterized by internal oscillations, propagate along the
chain. Typically the observed propagation velocities are {\it smaller} than
the maximum group velocity $max(|v_q|)= \sqrt{\phi_2}$.
Also the motion of these ILMs along the lattice leaves excited lattice
parts behind, implying that in the course of time these moving structures
will slow down or disappear due to radiation of energy.

While initially ILM excitations seemed to be connected
to some specific properties of FPU chains, the observation of
similar localized structures in Klein-Gordon chains as well as in higher dimensional
lattices suggested that the existence of ILMs is a rather
generic feature for various anharmonic Hamiltonian lattices \cite{DB-REVIEWS}.
This view became a well established fact due to numerous studies
during the past decade. In the following we will discuss the existence
and properties
of ILMs or DBs in FPU chains using methods of the modern theory of
localized excitations in discrete systems.

\section{Discrete breathers in FPU chains - some definitions}

Discrete breathers (intrinsic localized modes) are time-periodic
spatially localized solutions of the equations of motion of
a Hamiltonian lattice \cite{DB-REVIEWS}. 
In mathematical terms we are searching for solutions of
(\ref{1-2}) satisfying
\begin{eqnarray}
\hat{x}_l(t+T_b)=\hat{x}_l(t)\;,\;\hat{p}_l(t+T_b)=\hat{p}_l(t)\;,
\label{2-1} \\
\hat{x}_{l \rightarrow \pm \infty} \rightarrow d_{\pm}\;,\;
\hat{p}_{l \rightarrow \pm \infty} \rightarrow 0\;.
\label{2-2}
\end{eqnarray}
The difference $d_+ - d_-$ characterizes the DB induced lattice
deformation and is related to the abovementioned effects of
thermal expansion (or contraction). 
Without loss of generality we may choose $d_-=0$ in the following.
The period $T_b$ is
related to the DB frequency $\Omega_b=2\pi/T_b$.
The DB solution can be thus represented as a Fourier series expansion
with respect to time:
\begin{equation}
\hat{x}_l(t) = \sum_{k=-\infty}^{+\infty} A_{kl}{\rm e}^{ik\Omega_b t}\;.
\label{2-2b}
\end{equation}
The localization property (\ref{2-2}) implies 
\begin{equation}
A_{k\neq0,l\rightarrow \pm \infty} \rightarrow 0\;,\;
A_{k=0,l \rightarrow \pm \infty} \rightarrow d_{\pm}\;.
\label{2-2c}
\end{equation}
Inserting (\ref{2-2b}) into the equations of motion (\ref{1-2}),
and assuming a large distance from the DB core, the linearization
of the algebraic equations for the coefficients $A_{kl}$ together
with the condition (\ref{2-2c}) leads to
the {\it nonresonance}
condition \cite{sf94}
\begin{equation}
k\Omega_b \neq \omega_q
\label{2-2a}
\end{equation} for all integer $k$. Excluding $k=0$ for a moment,
that condition implies $\Omega_b > \omega_{\pi}$. The possible
resonance for $k=0$ which induces nonzero static lattice deformations,
will be discussed in more detail below.

The stability of a DB (as for any periodic orbit)
can be accounted for by linearizing the phase space flow around
a given DB solution $\hat{x}_l(t)$, i.e. by adding a small perturbation
to it $x_l(t)=\hat{x}_l(t) + \epsilon_l(t)$, inserting this
expression into the equations of motion (\ref{1-2}) and keeping only terms
linear in $\epsilon_l$:
\begin{eqnarray}
\dot{\epsilon}_l=\pi_l\;,\;
\nonumber
\dot{\pi}_l = -W''(\hat{x}_l-\hat{x}_{l-1})
(\epsilon_l - \epsilon_{l-1}) +
\\
W''(\hat{x}_{l+1}-\hat{x}_l)
(\epsilon_{l+1} - \epsilon_l)
\;.
\label{2-3}
\end{eqnarray}
Equations (\ref{2-3}) define a map
\begin{equation}
\left( 
\begin{array}{c}
\vec{\pi} (T_b) \\ \vec{\epsilon} (T_b) 
\end{array}
\right) = {\cal F} 
\left( 
\begin{array}{c}
\vec{\pi} (0) \\ \vec{\epsilon} (0) 
\end{array}
\right)
\label{2-4}
\end{equation}
which maps the phase space of perturbations onto itself
by integrating each point over the DB period $T_b$.
Here we used the abbreviation $\vec{x}\equiv (x_1,x_2,...,x_l,...)$.
The map (\ref{2-4}) is characterized by a symplectic Floquet matrix ${\cal F}$,
whose complex eigenvalues $\lambda$ and eigenvectors $\vec{y}$ provide
information about the stability of the DB. For details we
refer to \cite{sf04}. 
Here we note that if all eigenvalues $\lambda$
are of length one, then the DB is linearly (marginally) {\it stable}.
Otherwise perturbations exist which will grow in time (typically
exponentially) and correspond to a linearly {\it unstable} DB.
Upon changing a control parameter (e.g. the DB frequency) stable DBs
can become unstable (and vice versa). Such a change of stability is
appearing because two (or more) Floquet eigenvalues collide on the unit
circle and depart from it. If one of the two associated eigenvectors (or both)
is spatially localized, such a collision and the corresponding instability
are independent of the size of the lattice. If both eigenvectors are spatially
delocalized, the strength of the instability depends on the size of the system
and vanishes in the limit of an infinite system \cite{ma98}.

The absence of a cubic term $\phi_3$
(or more generally the case of a symmetric
potential $W$) implies a parity symmetry of the interaction
potential $W$. Consequently DB solutions will contain only odd harmonics
in a Fourier expansion with respect to time, $A_{2k,l}=0$. That implies $d_+=0$
(we remind the reader that $d_-=0$ as well).
For such a case the Floquet problem (\ref{2-3}) is periodic with period
$T_b/2$, because the time-periodic coefficients in (\ref{2-3}) contain only
even harmonics $2k\Omega_b$ in a Fourier expansion with respect to time.

Two eigenvectors of the Floquet matrix are corresponding to homogeneous
shifts in all coordinates or likewise velocities (due to the conservation of $P$).
Their eigenvalues are always located at +1 in the complex plane.
Two more eigenvectors of the Floquet matrix are corresponding to perturbations
along the DB periodic orbit (phase mode) or along the family of DB solutions.
Their eigenvalues are located either at +1 for asymmetric potentials $W$ (since
the Floquet mapping is performed for one period $T_b$) or at -1 for
symmetric potentials $W$ (since the Floquet mapping is performed for
$T_b/2$).

\begin{figure}[htb]
\vspace{20pt} \centering
\includegraphics[width=0.95\columnwidth]{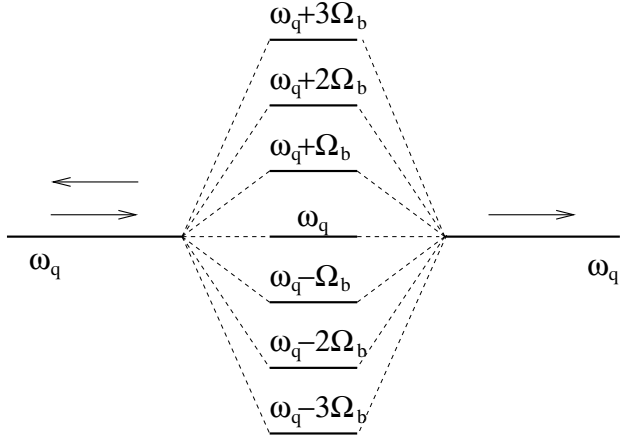}\\
\caption{Schematic representation of the one-channel scattering of a wave
by a discrete breather.}
\label{figscat}
\end{figure}

Finally we will also use the linearized phase space flow dynamics
around a DB solution (\ref{2-3}) in order to compute the
transmission coefficient for a small amplitude plane wave launched
into the DB.
It is a special Bloch-type solution of the equations (\ref{2-3}) of the form
\cite{caf98}
\begin{equation}
\epsilon_l(t) = \sum_{k=-\infty}^{\infty} \epsilon_{lk} {\rm e}^{i(\omega_q+k\Omega_b)t}
\;.
\label{bloch}
\end{equation}
The DB solution acts as a time-periodic scattering potential for an incoming wave
with frequency $\omega_q$ and generates new {\it channels} at frequencies
$\omega_q+k\Omega_b$. If $\omega_q+k\Omega_b \neq \omega_{q'}$ for nonzero $k$
then all channels are {\it closed} except for the open channel $k=0$.
Such a situation corresponds to {\it elastic}
scattering, i.e. the energies of the incoming and the outgoing (transmitted and reflected)
waves are equal. Otherwise we are confronted with {\it inelastic} scattering.
The scattering setup is schematically represented in Fig.\ref{figscat}.
%
For computational details the reader is referred to
\cite{fmf03}.

\section{Existence proofs}

The pioneering DB existence proof of MacKay and Aubry \cite{ma94}
was designed to apply for networks of weakly coupled oscillators
using the implicit function theorem. The extension of this technique
to FPU systems turned out to be complicated because one needs
a limiting case where DB solutions are compact.
Nevertheless it was applied successfully to the case of a {\it diatomic}
FPU chain with alternating heavy and light masses \cite{lsm97}. DB solutions were
shown to exist close to the limit where the ratio of light to heavy masses
vanishes.

A special case of {\it homogeneous} potentials $W$ with $\phi_{m}=
\mu \delta_{m,2n}$ where $n$ is a given positive integer,
was treated in \cite{ysk93} to separate time and space dependence.
This feature was then used in \cite{sf95} to provide
a constructive proof of existence of DBs. As shown recently \cite{fdmf03}, this
proof can be extended to systems with onsite potentials by
adding $\sum_l (\frac{1}{2}\omega_0^2 x_l^2 + \nu x_l^{2n})$ to the
Hamiltonian (\ref{1-1}).

An implicit proof of existence of DBs in FPU models was
provided by Aubry et al in \cite{akk01} using a variational method.
Discrete
breathers are obtained as loops in phase space which maximize a certain
average energy function for a fixed pseudoaction. More recently
James \cite{gj01} proved existence of DBs in FPU models for
low amplitudes (energies).

All these results provide the certainty of existence of
DB solutions in infinite FPU models. However, due to the implicit character
of most of these proofs a detailed study of DB properties
has to be obtained e.g. using advanced computational methods.

\section{DB solutions and their stability}

Computational tools for studying DB properties are confined to the case of a 
finite lattice size. The typically exponential spatial
degree of localization of DBs yields reliable results
which apply for infinite lattice size as well.
In terms of the dynamics in phase space $\{ x_l,p_l\}$ we are searching for
a starting point of a trajectory such that after integrating
over a given period of time the trajectory reaches the
starting point again. Of course all points on the corresponding
one-dimensional manifold generated by the trajectory fulfill this
condition. Discrete breathers are so-called isolated periodic orbits.
That implies that fixing the total energy and mechanical momentum
for a given DB solution, no other DB solutions are found in
an infinitesimal neighbourhood of the DB loop. In other words,
DB solutions do not belong to higher dimensional {\it resonant} tori
in the model phase space. However if changes in the energy are allowed,
then generically infinitesimal deformations of the original DB loop
will result in a new DB loop with slightly changed parameters
such as energy, frequency or amplitude. 
Sliding along such a family of DB solutions may lead to a stability
change of the solutions. Here we use a Newton method to find a DB
solution at a chosen frequency together with open boundary conditions. 
For details on this method as well as
on numerical evaluations of the Floquet matrix we refer to
\cite{sf04}. 

\subsection{The Page mode for $\phi_3=0$}

\begin{figure}[htb]
\vspace{20pt} \centering
\includegraphics[angle=270, width=0.95\columnwidth]{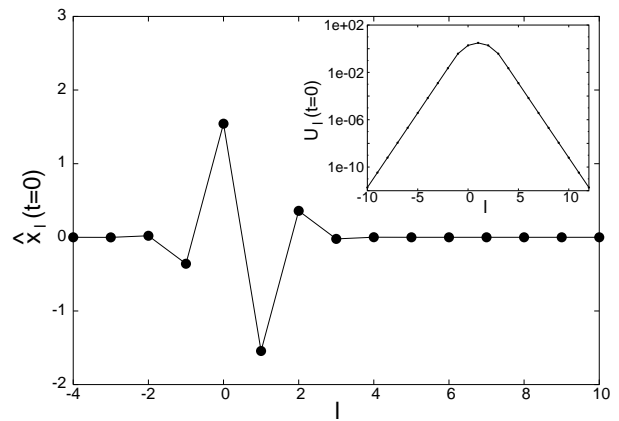}\\
\caption{Displacements
$\hat{x}_l(t=0)$ 
for $\phi_2=\phi_4=1$, $\phi_3=0$ and $\Omega_b=4.5$ versus lattice
site number $l$ for an antisymmetric DB.
Inset: Staggered deformation $u_l(t=0)$ for the same solution
on a logarithmic scale versus lattice site number $l$.
This DB is stable.}
\label{fig2}
\end{figure}

In Fig.\ref{fig2} we show the profile of a {\it stable} DB solution
$\hat{x}_l(t=0)$
for $\phi_2=\phi_4=1$, $\phi_3=0$, $\Omega_b=4.5$ and 80 sites,
the index $l$ running from $l=-39$ to $l=40$ with open boundary conditions. 
Note that $\hat{p}_l(t=0)=0$. Because $\Omega_b > \omega_{\pi}$,
the amplitudes stagger \cite{sf94}. 
%
The solution is antisymmetric in space, and likewise coined
{\it centered between sites} and also {\it Page mode}.
In the inset we plot the profile
of the corresponding {\it staggered deformation}
$u_l= (-1)^l (x_l - x_{l-1})$ on a
logarithmic scale. The DB is strongly localized,
and only a part of the chain is shown. The inset shows that
the DB is localized exponentially. The exponent depends
on the model parameters and the DB frequency \cite{sf94}.
%
\begin{figure}[htb]
\vspace{20pt} \centering
\includegraphics[angle=270,width=0.95\columnwidth]{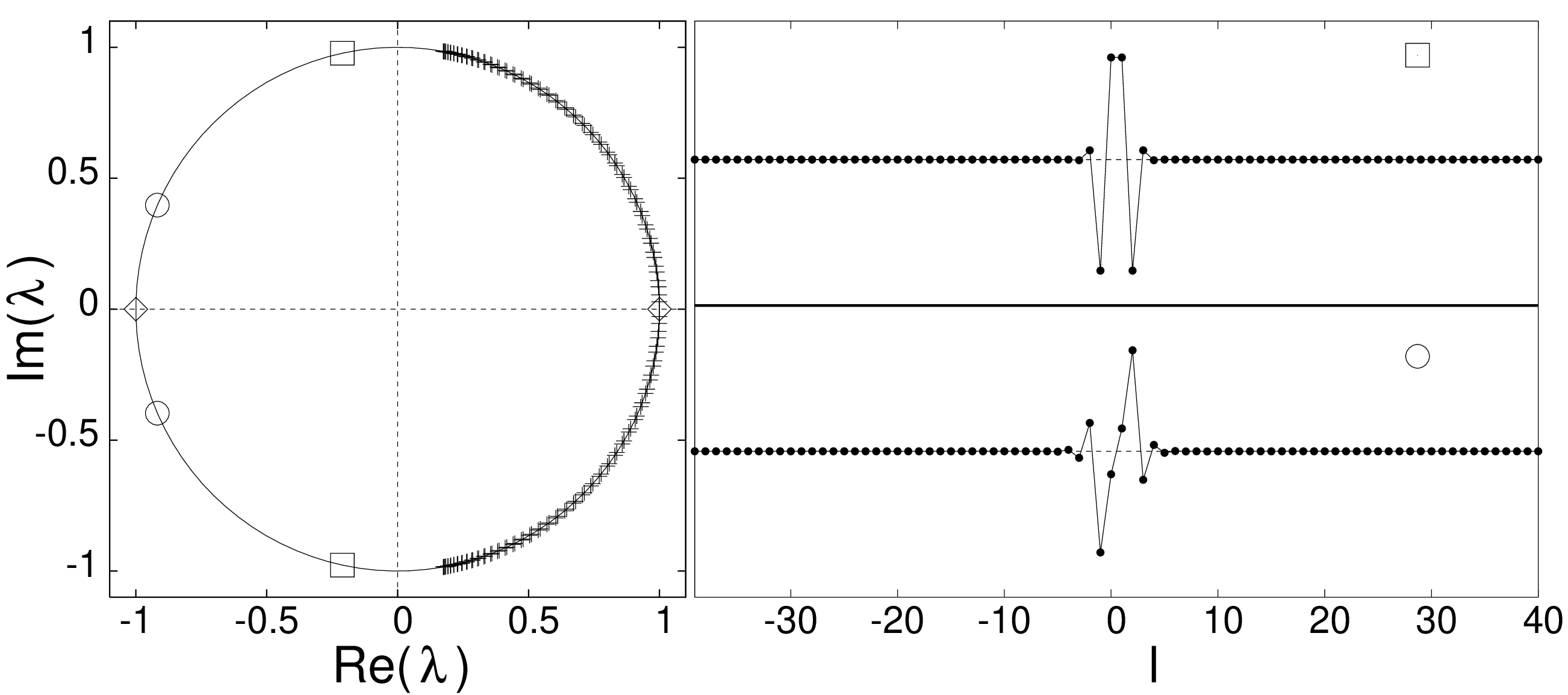}\\
\caption{
Left panel: location of Floquet eigenvalues $\lambda$ in the complex
plane for the
DB in Fig.\ref{fig2} (crosses, diamonds, squares, circles). 
The unit circle is shown to guide the eye.
Right panels: real part of the displacement components of
the Floquet eigenvectors marked with the corresponding symbols
(square and circle). }
\label{fig2-2}
\end{figure}
In the left part of Fig.\ref{fig2-2} we show the location of
the Floquet eigenvalues $\lambda$ in the complex plane.
All eigenvalues reside on the unit circle. The two pairs
of degenerate eigenvalues located at +1 and -1 on the real axis
(see discussion above) are shown with diamonds. 
Besides the Floquet continuum of extended eigenstates (crosses) two pairs
of stable spatially localized eigenstates are observed
(shown by squares and circles). 
In the two right panels of Fig.\ref{fig2-2} the real part of
the displacement components of these
localized eigenstates are plotted.

\subsection{The Sievers-Takeno mode for $\phi_3=0$}

\begin{figure}[htb]
\vspace{20pt} \centering
\includegraphics[angle=270, width=0.95\columnwidth]{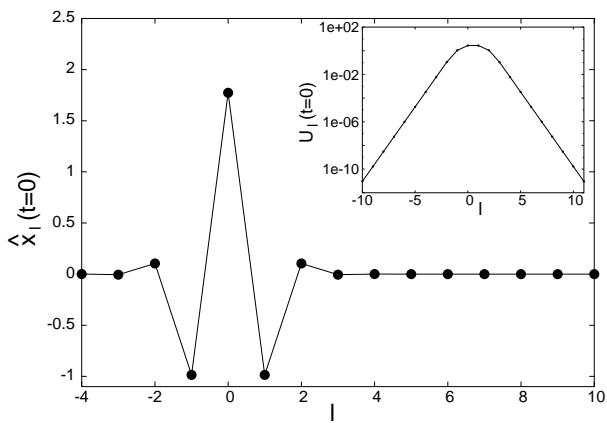}\\
\caption{Displacements
$\hat{x}_l(t=0)$
for $\phi_2=\phi_4=1$, $\phi_3=0$ and $\Omega_b=4.5$ versus lattice
site number $l$ for a symmetric DB.
Inset: Staggered deformation $u_l(t=0)$ for the same solution
on a logarithmic scale versus lattice site number $l$.
This DB is unstable.}
\label{fig3}
\end{figure}

For the same parameters we show in 
Fig.\ref{fig3} an {\it unstable} DB solution which is symmetric
in space, or {\it centered on a site} and also known as the
{\it Sievers-Takeno mode}.
%

\begin{figure}[htb]
\vspace{20pt} \centering
\includegraphics[angle=270,width=0.95\columnwidth]{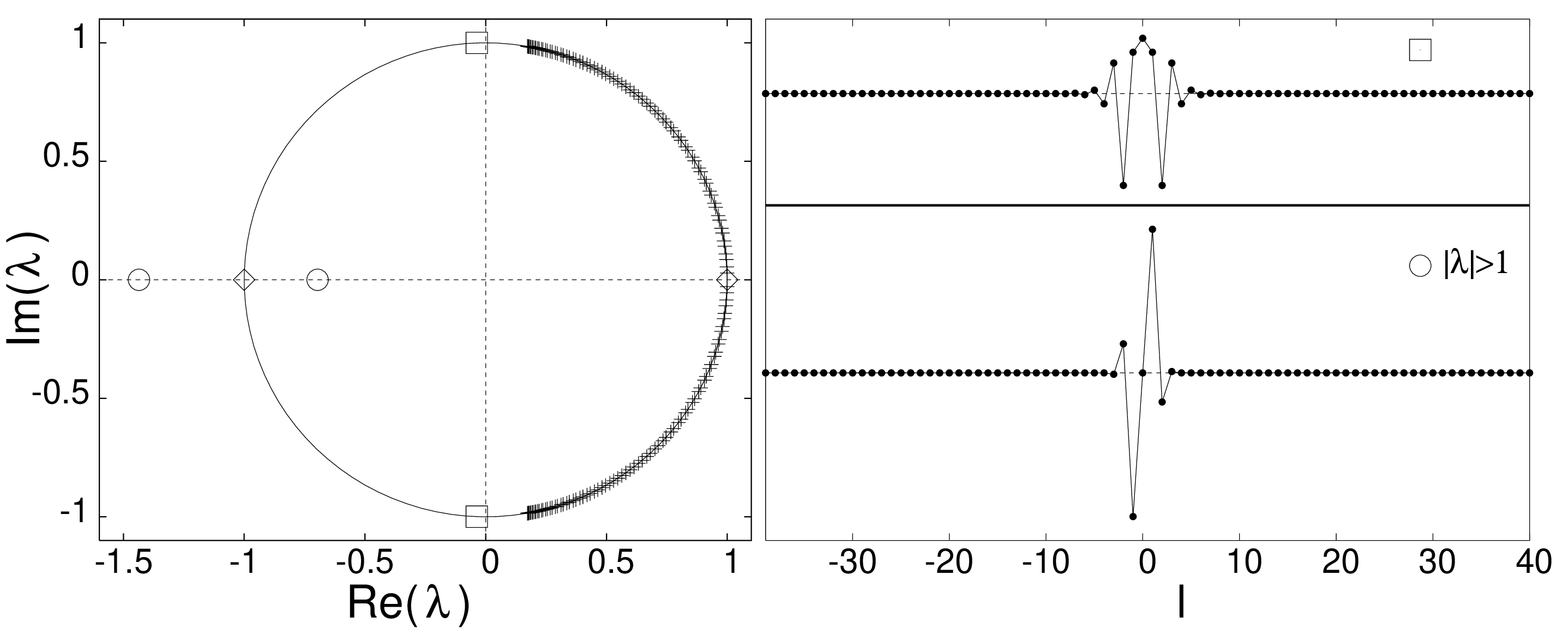}\\
\caption{Left panel: location of Floquet eigenvalues $\lambda$ in the complex
plane for the
DB in Fig.\ref{fig3} (crosses, diamonds, squares, circles). 
The unit circle is shown to guide the eye.
Right panels: real part of the displacement components of
the Floquet eigenvectors marked with the corresponding symbols
(square and circle).}
\label{fig3-2}
\end{figure}

In the left part of Fig.\ref{fig3-2} we show the location of
the Floquet eigenvalues $\lambda$ in the complex plane.
Not all eigenvalues reside on the unit circle.
At variance with the Page mode,
one of the pairs of localized eigenstates (circles)
is located on the real axis off the unit circle.
%
In the two right panels of Fig.\ref{fig2-2} the real part of
the displacement components of these
localized eigenstates are plotted.
The instability is caused by a spatially localized Floquet eigenvector,
which deforms the Sievers-Takeno mode in Fig.\ref{fig3} in the direction
of the Page mode in Fig.\ref{fig2}.
Indeed the Sievers-Takeno mode, when perturbed in the full nonlinear equations
of motion along the unstable Floquet eigenvector, starts to perform additional 
oscillations around the stable Page mode for times large compared to the
DB period.

\subsection{The Page mode for $\phi_3\neq 0$}

\begin{figure}[htb]
\vspace{20pt} \centering
\includegraphics[angle=270,width=0.95\columnwidth]{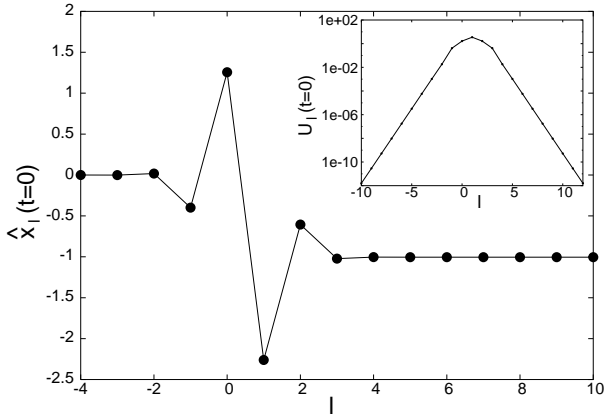}\\
\caption{Displacements $\hat{x}_l(t=0)$
for $\phi_2=\phi_3=\phi_4=1$ and $\Omega_b=4.5$ versus lattice
site number $l$.
Inset: Staggered deformation $u_l(t=0)$ for the same solution
on a logarithmic scale versus lattice site number $l$.
This DB is unstable.}
\label{fig4}
\end{figure}

If $\phi_3 \neq 0$, then the interaction potential $W$ becomes asymmetric.
DB solutions will now contain in general all Fourier harmonics $k\Omega_b$
including $k=0$.
Depending on the sign of $\phi_3$ this will cause either a
contraction or expansion of the chain. In Fig.\ref{fig4} we show
a DB solution which can be continued from the Page mode in Fig.\ref{fig2}
up to $\phi_3=1$.
%
For that particular case $d_+\approx -1.005$. The DB solution can be thus
viewed as a localized vibration which induces a nonzero kink-shaped
lattice distortion. This DB solution turns out to be {\it unstable}.
The reason for that is the interaction of {\it localized} Floquet eigenstates
with the Floquet continuum (the eigenstates which correspond to
plane waves far away from the DB). Such {\it oscillatory instabilities}
can be characterized by the wave numbers $q_{os}$ which correspond
to the extended Floquet state interacting with the originally localized one \cite{ma98}.
%
\begin{figure}[htb]
\vspace{20pt} \centering
\includegraphics[angle=270, width=0.95\columnwidth]{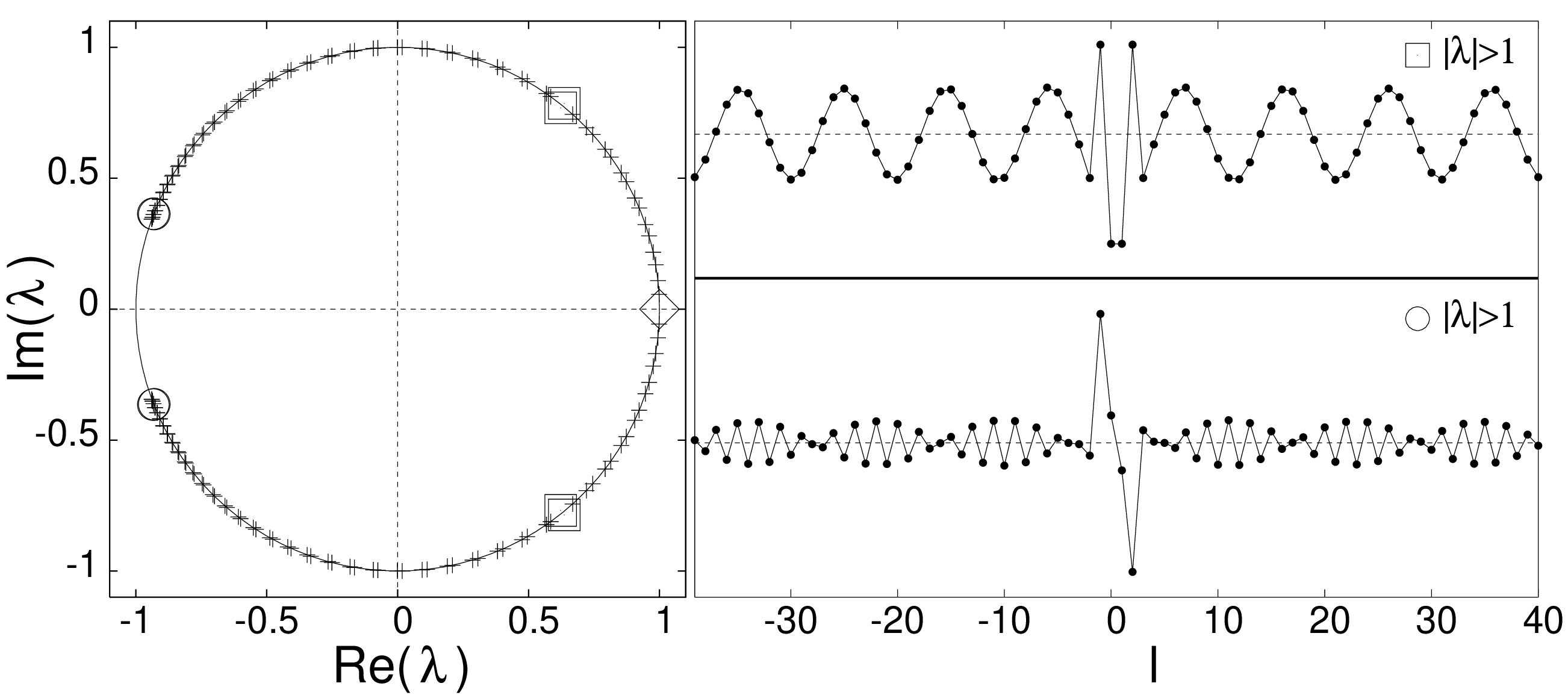}\\
\caption{Left panel: location of Floquet eigenvalues $\lambda$ in the complex
plane for the
DB in Fig.\ref{fig4} (crosses, diamonds, squares, circles).
The unit circle is shown to guide the eye.
Right panels: real part of the displacement components of
the Floquet eigenvectors marked with the corresponding symbols
(square and circle).}
\label{fig4-2}
\end{figure}
In the left part of Fig.\ref{fig4-2} we show the location of
the Floquet eigenvalues $\lambda$ in the complex plane.
Not all eigenvalues reside on the unit circle. The two pairs
of degenerate eigenvalues, which were located at +1 and -1 on the real axis
(see discussion above) for $\phi_3=0$, are now all located
at +1 and shown with diamonds. 
Besides the Floquet continuum of extended eigenstates two pairs
of eigenstates (which were localized for the Page mode with $\phi_3=0$) are observed
to strongly interact with the Floquet continuum
(shown by squares and circles). While the small deviation of
the corresponding eigenvalues from the unit circle
is of the order of $0.01 ... 0.03$, we checked that it is independent
of the system size by increasing the number of sites from 80 to 640 (see also \cite{ma98}).
In the two right panels of Fig.\ref{fig4-2} the real part of
the displacement components of these resonating and unstable
eigenstates are plotted.

When the Page mode is perturbed along the oscillatory instability eigenvector
in the full nonlinear equations of motion, it deforms initially very slowly
compared to the DB period, due to the weak amplitude of the instability.
After reaching a critical threshold in the perturbation amplitude,
the mode depins from its original lattice site position and starts to move
along the lattice.

\subsection{ The Sievers-Takeno mode for $\phi_3 \neq 0$}

When continuing the Sievers-Takeno mode from $\phi_3=0$ to nonzero values
of $\phi_3$, the strong instability caused by a localized Floquet eigenstate
for $\phi_3=0$ remains. It also acquires the above discussed oscillatory
instabilities of the Page mode. At the same time the spatial profile of the mode
ceases to show up with any symmetry.

\subsection{DB stability properties for different frequencies}

In the following we present results on the stability of the Page mode DB
upon variation of the DB frequency $\Omega_b$. The Sievers-Takeno mode DB
yields similar results, except for keeping its strong localized instability
discussed above. 

%
\begin{figure}[htb]
\vspace{20pt} \centering
\includegraphics[angle=270,width=0.95\columnwidth]{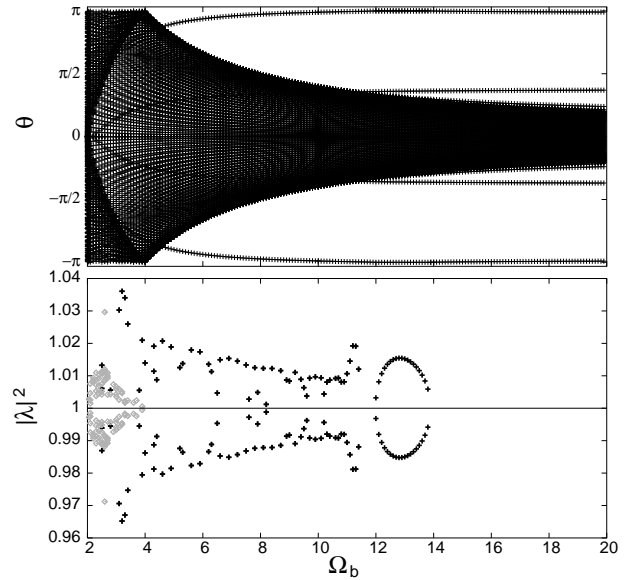}\\
\caption{All Floquet phases $\theta$ and the squared eigenvalue length
$|\lambda|^2$ of the unstable Floquet eigenvectors
for a DB with $\phi_2=\phi_4=1$ and $\phi_3=0.5$
as a function of the DB frequency $\Omega_b$.
Eigenvalues which correspond to finite size instabilities are shown using
grey symbols.}
\label{fig5}
\end{figure}
In Fig.\ref{fig5} we plot
the dependence of the arguments $\theta$ and the squared absolute values
$|\lambda|^2$ of the irreducible part of the Floquet matrix eigenvalue
spectrum $\lambda = |\lambda| {\rm exp}(i\theta)$ as a function of
the DB frequency $\Omega_b$ for $\phi_2=\phi_4=1$ and $\phi_3=0.5$.
These results
are obtained for a DB family which is continued from
the Page mode in Fig.\ref{fig2}.
We first note that 
$\omega_{\pi}\equiv2$ and for $\omega_{\pi}
< \Omega_b < 2\omega_{\pi}$ the arguments $\theta$ cover the
whole unit circle interval. The corresponding interactions between
different extended Floquet eigenstates lead to finite size instabilities
\cite{ma98},\cite{akk01}
which are marked with grey symbols in the lower panel of Fig.\ref{fig5}.
These instabilities will disappear for an infinite system, i.e. the
corresponding values $|\lambda| - 1 \sim 1/N$ where $N$ is the number of
sites \cite{ma98}. For that DB frequency interval scattering of
waves will be inelastic for wavenumbers $q,q'$ satisfying
$\omega_q + \Omega_b = \omega_{q'}$ \cite{caf98}. Note that this inelastic
two-channel scattering will survive in the limit of an infinite system,
opposite to the finite size instabilities of the DB itself.

The mentioned oscillatory instabilities induced by the
interaction between two pairs of originally localized eigenstates and the continuum
are clearly observable, and cause deviations of the corresponding
absolute values $|\lambda|$ from unity. 
As conjectured in \cite{sjca04}, these instabilities survive in the limit
of an infinite lattice.
The first pair of these eigenstates exits the continuum at $\Omega_b \approx 4.6$.
The corresponding eigenvalues return to the unit circle and 
approach $-1$ in the complex plane at $\Omega_b \approx 12$,
collide with each other, separate on the real axis causing another instability (cf.  the 
corresponding ellipse structure in Fig.\ref{fig5}) and merge again at -1 for $\Omega_b \approx
14$. 
The second pair exits the continuum at $\Omega_b \approx 11.5$.
For sufficiently large values
of $\Omega_b > 14$ the DB becomes {\it stable}. 

The discussed instabilities for the Page mode are solely caused by the asymmetric term $\phi_3$.
Indeed, for $\phi_3=0$ the abovementioned periodicity of (\ref{2-3}) with
$T_b/2$ excludes the possibility of finite size instabilities or equally inelastic
multichannel scattering, since the condition 
$\omega_q + 2\Omega_b = \omega_{q'}$ can not be satisfied due to $\Omega_b > \omega_{\pi}$.
%
\begin{figure}[htb]
\vspace{20pt} \centering
\includegraphics[angle=270,width=0.95\columnwidth]{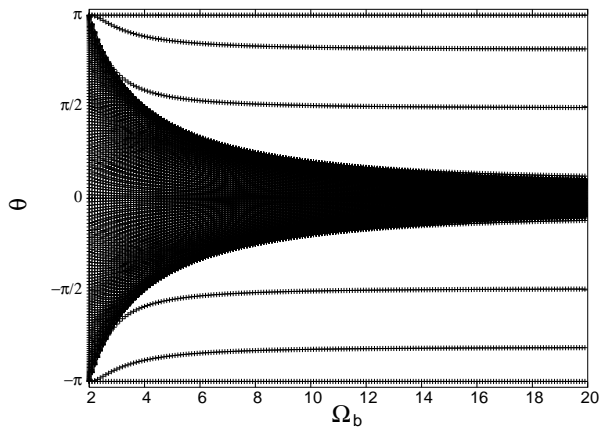}\\
\caption{All Floquet phases $\theta$ 
for a DB with $\phi_2=\phi_4=1$ and $\phi_3=0$
as a function of the DB frequency $\Omega_b$. Note that the Floquet map is obtained 
here by integrating (\ref{2-3}) over the time $T_b/2$.}
\label{fig6}
\end{figure}
The oscillatory instabilities caused by the 
interaction of localized Floquet eigenstates with the Floquet
continuum are removed for the same reason. The dependence of the eigenvalue phases $\Theta$
on $\Omega_b$ for a Floquet matrix defined by integrating (\ref{2-3}) over $T_b$  
for $\phi_3=0$ is practically identical to the
case $\phi_3=0.5$
(upper panel in Fig.\ref{fig5}).  
However the correct irreducible Floquet matrix for $\phi_3=0$ is obtained
by integrating (\ref{2-3}) over the true period $T_b/2$.
The phase variation 
shown in  Fig.\ref{fig6} demonstrates that the
localized Floquet states are indeed separated from the Floquet continuum. 

\section{Resonant wave scattering by DBs}

As mentioned above, the linearized phase space flow equations (\ref{2-3}) give
also information about the complex transmission amplitude $t_q$ and the transmission
coefficient $T(q) \equiv |t_q|^2$. While the numerical technique and also a number of results
concerning wave scattering by FPU breathers have been reported recently
\cite{caf98},\cite{fmf03},\cite{fmff03}, we want
to address here one aspect which was not yet discussed. 
%
\begin{figure}[htb]
\vspace{20pt} \centering
\includegraphics[angle=270,width=0.95\columnwidth]{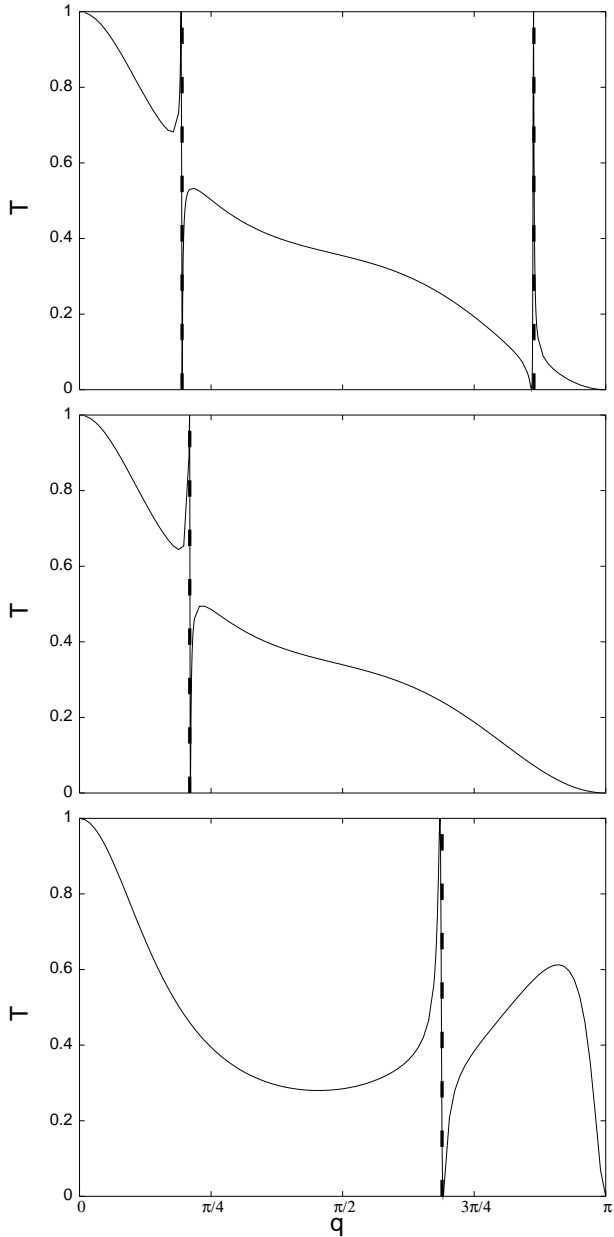}\\
\caption{
Transmission coefficient $T$ versus wave number $q$ of an incident wave
for $\phi_2=\phi_4=1$
and $\phi_3=0.5$ scattered by a Page mode DB. From top to bottom:
$\Omega_b=4.5,4.7,10$.
The vertical dashed line indicates the position of the oscillatory
instabilities $q_{os}$.}
\label{fig7}
\end{figure}

We show in Fig.\ref{fig7} the $q$-dependence of $T(q)$ for three different
frequencies $\Omega_b=4.5,4.7,10$ of the Page mode DB with $\phi_2=\phi_4=1$
and $\phi_3=0.5$, for which we gave the Floquet spectral data in Fig.\ref{fig5}.
The perfect transmission  $T(q=0)=1$ for all cases is due to the conservation of total
mechanical momentum \cite{fmf03}. Equally the vanishing of the transmission  $T(q=\pi)=0$ is
due to the vanishing of the group velocity (\ref{1-7}) at the band edge $q=\pi$
\cite{caf98}.
However the resonant perfect transmission and reflection peaks observed for
$q\neq 0,\pi$ are due to {\it Fano resonances} \cite{fmf03},\cite{fmff03}.  We remind the reader that
Fano resonances are induced by localized states interacting with a continuum
of scattering states \cite{uf61}.

\begin{figure}[htb]
\vspace{20pt} \centering
\includegraphics[angle=270,width=0.95\columnwidth]{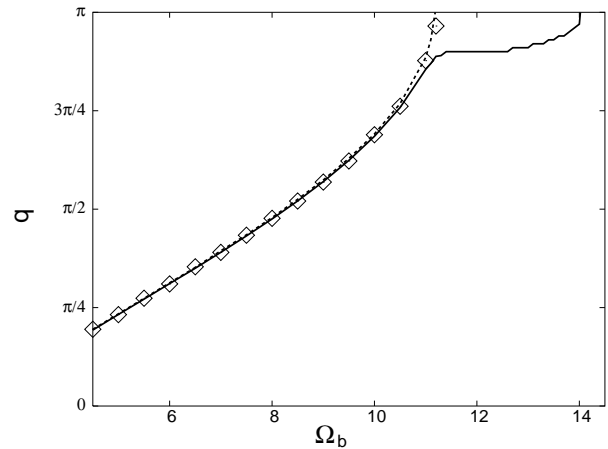}\\
\caption{
The dependence of the position of an oscillatory instability $q_{os}$
(diamonds)
and the associated locations of perfect transmission (solid line)
and perfect reflection (dashed line) on $\Omega_b$ for
a Page mode DB with $\phi_2=\phi_4=1$
and $\phi_3=0.5$.}
\label{fig8}
\end{figure}

This is precisely the situation observed in the case of an asymmetric FPU potential.
For $\phi_3=0$ the Page mode DB has a localized Floquet eigenstate well separated
from the Floquet continuum. However any nonzero value of $\phi_3$ induces
an interaction between this localized state and the continuum mediated by the DB.
This interaction causes an oscillatory instability in the Floquet matrix spectrum.
The dashed vertical lines in all three panels in Fig.\ref{fig7} indicate the location
of the oscillatory instability $q_{os}$ as observed
in Fig.\ref{fig5}. In all three cases the value of $q_{os}$ is very close to the
locations of perfect transmission and reflection. 
In Fig.\ref{fig8} we plot the $q$-values of
%
the oscillatory instability of the DB $q_{os}$ from Fig.\ref{fig5} together with the positions of
the perfect transmission and reflection as functions of $\Omega_b$. The obtained correlation
between these resonant $q$-values is evident.

From a general point of view, the DB acts as a time-periodic
scattering potential for incoming waves. The time-periodic scattering potential
generates 
locally new frequencies from an incoming wave which correspond to closed channels. 
These closed channels provide
additional propagation ways for the incoming wave. In that sense a time-periodic scattering
potential in a strictly one-dimensional system acts similar to a local enlargement of the
system dimensionality around the scatterer \cite{caf98},\cite{fmf03}.
Such additional propagation channels allow for wave interference, and may also cause
a total {\it destructive interference} leading to a total reflection of the incoming wave,
which is another way of understanding the origin of the observed Fano resonance \cite{fmff03}.
Note that in a strict sense the correlation between an oscillatory instability of a DB solution and a
corresponding Fano resonance holds only for the limit of weak coupling between the
localized Floquet state and the Floquet continuum \cite{fmf03}. And only in that limiting case we can
expect a nearby location of perfect reflection (Fano resonance) and perfect transmission.
The case of strong interaction can lead to the disappearance of an oscillatory instability in
the Floquet spectrum and of the perfect transmission peak, but simultaneously to
a remaining of the Fano resonance of perfect reflection \cite{fmff03}.

\section{The frequency limits of the DB solution families}

When the frequency $\Omega_b$ is varied along a DB family, we
may consider the two limiting cases $\Omega_b \rightarrow \infty$
and $\Omega_b \rightarrow \omega_{\pi}$. The high frequency limit
implies large amplitudes inside the DB core. The leading order contribution
in the equations of motion will then be due to the $\phi_4$ term.
The core profile of these high frequency DBs will be very similar
to the DB cores in Figs.\ref{fig2},\ref{fig3} regardless the values
of $\phi_2$ and $\phi_3$ which will only affect the tail characteristics
\cite{sf94},\cite{sf95},\cite{ma96},\cite{deft01}.
Opposite to that, the low frequency limit is depending on the value
of $\phi_3$ \cite{sjca04},\cite{mk04}. 
%
\begin{figure}[htb]
\vspace{20pt} \centering
\includegraphics[angle=270,width=0.95\columnwidth]{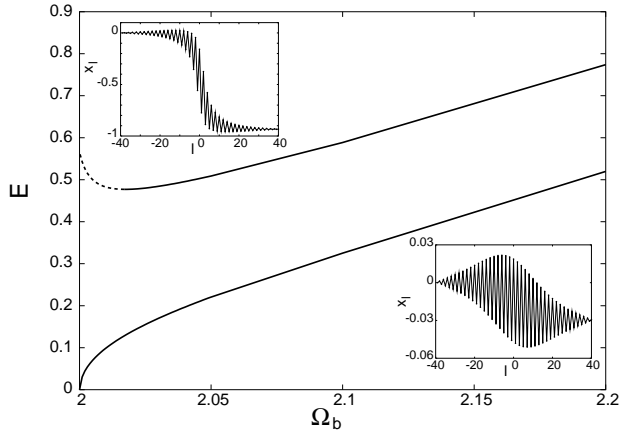}\\
\caption{
DB energy $E$ versus DB frequency $\Omega_b$ for $\phi_2=\phi_4=1$
and $\phi_3=0.5$ (lower curve) and $\phi_3=1$ (upper curve).
Insets:
Displacements $\hat{x}_l(t=0)$
for $\Omega_b=2.001$ versus lattice
site number $l$. Upper left inset for $\phi_3=1$,
lower right inset for $\phi_3=0.5$. }
\label{fig9}
\end{figure}

In Fig.\ref{fig9} we plot the dependence of the DB energy (\ref{1-1}) on
its frequency for $\phi_3=1$ and $\phi_3=0.5$ and 80 sites. For $\phi_3=0.5$
the energy tends to zero as $\Omega_b$ tends to $\omega_{\pi}$.
However for $\phi_3=1$ the DB energy passes through a minimum as the frequency
is lowered, and increases again with further approaching the limiting
value $\omega_{\pi}$ \cite{sjca04},\cite{mk04}. These results are intimately connected with
the modulational instability of plane waves at small amplitudes 
\cite{bb83},\cite{sp94},\cite{sf96}.
Assuming that a DB family has a zero amplitude limit, it follows
that in a one-dimensional chain as discussed here its energy will also
vanish in that limit \cite{fkm97}. It has been 
conjectured \cite{sf96} (and later confirmed numerically \cite{sp94})
that a small-amplitude DB occurs due to an instability (bifurcation)
of the $q=\pi$ plane wave. The rigorous analysis of this plane wave
bifurcation gives an inequality
\cite{bb83},\cite{sp94},\cite{sf96},\cite{gj01}
\begin{equation}
3\phi_2\phi_4 - 4\phi_3^2 > 0
\label{2-5}
\end{equation}
which has to hold in order to obtain the modulational instability
of the $q=\pi$ mode and by conjecture a low-amplitude DB solution.
Since inequality (\ref{2-5}) is satisfied for $\phi_2=\phi_4=1$ and $\phi_3=0.5$,
we do observe the low amplitude (and thus low energy) limit of the
DB family in Fig.\ref{fig9}. However choosing $\phi_2=\phi_4=1$
and $\phi_3=1$, the inequality (\ref{2-5}) is violated. Consequently the $q=\pi$ mode
does not experience a low amplitude modulational instability,
and low amplitude
(and low energy) DBs do not exist. Nevertheless, as explained above,
the high frequency limit of such an FPU model will always allow for DB solutions
of a well known asymptotic shape. Consequently the energy versus frequency
curve in Fig.\ref{fig9} has to show up with a minimum.
It is instructive to compute the DB profiles for $\Omega_b$ close to
$\omega_{\pi}$ for the two cases. These profiles are shown in the
insets in Fig.\ref{fig9}. For $\phi_3=0.5$ the low energy DB
delocalizes and takes the shape of a $q=\pi$ plane wave (or better to
say a standing
wave, due to the open boundary conditions).
The lattice deformation $d_+ \approx -0.03$ tends to zero when increasing
the system size and further approaching $\Omega_b \rightarrow 2$.
For $\phi_3=1$
the DB with frequency close to $\omega_{\pi}$ has still large amplitudes,
including a nonzero and well pronounced lattice distortion $d_+ \approx -1$.
While both DB branches have weak oscillatory instabilities as shown in Fig.\ref{fig5},
the dashed part of the upper curve in Fig.\ref{fig9} indicates
that DBs from that part of the branch experience an additional
instability with corresponding eigenvalues located on
the real axis. When decreasing $\phi_3$ and approaching
$\phi_3^{(cr)}=\sqrt{3\phi_2\phi_4}/2$, which changes the sign of the l.h.s. of inequality
(\ref{2-5}), the value of $d_+$ for DBs with frequencies close to the linear spectrum
($\Omega_b \rightarrow 2$) tends to zero  and the energy
threshold disappears.

\section{Discussion}

The results of this work can be continued - and have been continued - in many
different directions. The unstable Sievers-Takeno mode DB, when properly perturbed,
lead to DB-like excitations which propagate along the lattice 
\cite{followups},\cite{movingfpu}. Some perturbative
analytical results give a partial understanding of this effect \cite{movingpt}. Still
there is strong evidence of the {\it nonexistence}
of exact and spatially localized moving DBs \cite{fk99}.

The inequality (\ref{2-5}) can be used to predict the existence or noexistence of DB solutions
for various other FPU-like models. A consequence is that DB excitations are not expected to
exist for the Toda chain. A similar conclusion can be obtained for a so-called {\it Roto-FPU}
model with $W(x)= 1-\cos x$. However such a system of coupled rotors allows for {\it rotobreathers},
i.e. solutions where a few degrees of freedom are in a rotating state, while the rest of the system
is in a spatially localized oscillating state 
\cite{tp96} (cf. Fig.\ref{figbreather_types}(b)).

Discrete breathers are not an exclusive property of one-dimensional lattices. They exist equally in
higher dimensional lattices \cite{DB-REVIEWS}, 
where they show up with nonzero lower energy thresholds \cite{fkm97}, independent
on further model parameters. 
This is true also for generalizations of FPU models to higher lattice dimensions.
An interesting result concerns the case of asymmetric interaction potentials $W$ and higher-dimensional 
FPU lattices. As shown in \cite{fkt97}, the requirement to have a finite energy for a DB leads to
a static lattice deformation with a dipole symmetry and a spatial decay law $\sim r^{1-d}$ where $r$ is
the distance from the breather core and $d$ the lattice dimension.

Finally we mention a number of studies of FPU breathers emerging from perturbed extended states
via modulational instability or induced by fluctuations in thermal equilibrium
\cite{dkpr05},\cite{statistics}. These studies show
clearly that DB solutions are not only interesting mathematical objects, but essential in order to
understand and describe the properties of nonlinear lattice dynamics in thermal equilibrium and
during relaxational processes.
\\
\\
\\
Acknowledgements
\\
We thank V. Fleurov, F. Izrailev, M. Johansson, Yu. Kosevich, Yu. S. Kivshar, R.S. MacKay,
A. E. Miroshnichenko and A.J. Sievers
for useful discussions.


\begin{thebibliography}{10}

\bibitem{fpu55}
E. Fermi, J. Pasta and S. Ulam, Los Alamos Science Laboratory Report No. LA-1940 (1955),
unpublished; reprinted in Collected Papers of Enrico Fermi, edited by
E. Segre (University of Chicago Press, Chicago, 1965), Vol. 2, p. 978.
Also in {\it The Many-Body Problem}, C. C. Mattis (Ed.), World Scientific,
Singapore, 1993. 

\bibitem{ck96}
C. Kittel, Introduction to solid state physics, Wiley, New York, 1996.

\bibitem{ic66}
F. M. Izrailev and B. V. Chirikov, Sov. Phys. Dokl. {\bf 11}, 30 (1966).

\bibitem{DB-REVIEWS}
A. J. Sievers and J. B. Page, in:
Dynamical properties of solids VII phonon physics the cutting edge,
Elsevier, Amsterdam (1995);
S. Aubry, Physica D {\bf 103}, 201 (1997);
S. Flach, C.R. Willis, Phys. Rep. {\bf 295}, 181 (1998);
Energy Localisation and Transfer, Eds. T. Dauxois,
A. Litvak-Hinenzon, R. MacKay and A. Spanoudaki,
World Scientific (2004);
D. K. Campbell, S. Flach and Yu. S. Kivshar,
Physics Today 57, 43 January (2004).

\bibitem{binder00a} P. Binder, D. Abraimov, A. V. Ustinov, S. Flach, 
and Y. Zolotaryuk,
{Phys. Rev. Lett.} {\bf 84}, 745 (2000); E. Trias, J. J. Mazo, 
and T. P. Orlando,
{Phys. Rev. Lett.} {\bf 84}, 741 (2000); A. Ustinov. 
{Chaos} {\bf 13}, 716 (2003).

\bibitem{eisenberg98} H. S. Eisenberg {\em et al.}, 
{Phys. Rev. Lett.} {\bf 81}, 3383 (1998);
 R. Morandotti {et al.},  {Phys. Rev. Lett.} 
 {\bf 83}, 2726 (1999);  {\bf 83}, 4756
(1999); J. W. Fleischer, M. Segev, N. K. Efremidis, and D. N.
Christodoulides, {Nature} {\bf 422}, 147 (2003);
D.Cheskis et.al., Phys. Rev. Lett. {\bf 91}, 223901 (2003).

\bibitem{swanson99} B. Swanson {et al.},  {Phys. Rev.
Lett.} {\bf 82}, 3288 (1999);
K. Kladko, J. Malek and A. R. Bishop, {J. Phys.: Condens. Matter}
{\bf 11}, L415 (1999).

\bibitem{schwarz99} U.T. Schwarz, L.Q. English, and A.J. Sievers,
{Phys. Rev. Lett.} {\bf 83}, 223
(1999).

\bibitem{sato03} M. Sato, B. E. Hubbard, A. J. Sievers, B. Ilic, D. A.
Czaplewski, and H. G. Craighead, {Phys. Rev. Lett.} {\bf 90}, 044102
(2003); M. Sato, B. E. Hubbard, A. J. Sievers  et al.,
Europhys. Lett. {\bf 66}, 318 (2004).

\bibitem{BEC}
B. Eiermann, Th. Anker, M. Albiez, M. Taglieber, P. Treutlein,
K.-P. Marzlin and M. K. Oberthaler, Phys. Rev. Lett. {\bf 92},
230401 (2004).

\bibitem{hightc}
M. Machida and T. Koyama, Phys. Rev. B {\bf 70}, 024523 (2004).

\bibitem{plasma}
I. Kourakis and P. K. Shukla, Phys. Plasmas {\bf 12}, 014502 (2005).

\bibitem{bb83}
N. Budinsky and T. Bountis, Physica D {\bf 8}, 445 (1983).

\bibitem{kp94}
Yu. S. Kivshar and M. Peyrard, Phys. Rev. A {\bf 46}, 3198 (1992).

\bibitem{sp94}
K. W. Sandusky and J. B. Page, Phys. Rev. B {\bf 50}, 866 (1994).

\bibitem{sf96}
S. Flach, Physica D {\bf 91}, 223 (1996).

\bibitem{pr97}
P. Poggi and S. Ruffo, Physica D {\bf 103}, 251 (1997).

\bibitem{dkpr05}
T. Dauxois, R. Khomeriki, F. Piazza and S. Ruffo,
this issue.

\bibitem{st88}
A. J. Sievers and S. Takeno, Phys. Rev. Lett.
{\bf 61}, 970 (1988).
%
S. Takeno and A.J. Sievers, Solid State Comm. {\bf 67}, 1023 (1988).
%
S. Takeno, K. Kisoda and A. J. Sievers, Prog. Theor. Phys.
{\bf 94} (Suppl), 242 (1988).

\bibitem{followups}
S. Takeno, J. Phys. Soc. Japan {\bf 59}, 3127 (1990);
S. Takeno, J. Phys. Soc. Japan {\bf 59}, 3861 (1990);
S. Takeno and K. Hori, J. Phys. Soc. Japan {\bf 59}, 3037 (1990);
S. Takeno and S. Homma, J. Phys. Soc. Japan {\bf 59}, 1890 (1990);
J. B. Page, Phys. Rev. B {\bf 41}, 7835 (1990);
V. M. Burlakov, S. A. Kisilev and V. I. Rupasov, JETP Lett. {\bf 51},
544 (1990);
V. M. Burlakov, S. A. Kisilev and V. I. Rupasov, Phys. Lett. A
{\bf 147}, 130 (1990);
V. M. Burlakov, S. A. Kisilev and V. N. Pyrkov,
Phys. Rev. B {\bf 42}, 4921 (1990);
V. M. Burlakov, S. A. Kisilev and V. N. Pyrkov,
Solid State Comm. {\bf 74}, 327 (1990);
S. A. Kisilev, Phys. Lett. A {\bf 148}, 95 (1990);
S. A. Kisilev and V. I. Rupasov, Phys. Lett. A {\bf 148}, 355 (1990);
S. Takeno and S. Homma, J. Phys. Soc. Japan {\bf 60}, 731 (1991);
S. Takeno and K. Hori, J. Phys. Soc. Japan {\bf 60}, 947 (1991);
V. M. Burlakov and S. A. Kisilev, JETP {\bf 72}, 854 (1991);
K.W. Sandusky, J.B. Page, and K.E. Schmidt, Phys. Rev. B {\bf 46},
6161 (1992);
S. Takeno and S. Homma, J. Phys. Soc. Japan {\bf 62}, 835 (1993);
S. R. Bickham, S. A. Kisilev and A. J. Sievers,
Phys. Rev. B {\bf 47}, 14206 (1993);
O. Chubykalo and Yu. S. Kivshar, Phys. Lett. A {\bf 178}, 123 (1993).

\bibitem{sf94}
S. Flach, Phys. Rev. E {\bf 50}, 3134 (1994).

\bibitem{sf04}
S. Flach, in: Energy Localisation and Transfer, Eds. T. Dauxois,
A. Litvak-Hinenzon, R. MacKay and A. Spanoudaki,
World Scientific (2004).

\bibitem{ma98}
J. L. Marin and S. Aubry, Physica D {\bf 119}, 163 (1998);
M. Johansson and Yu. S. Kivshar, Phys. Rev. Lett. {\bf 82}, 85 (1999).

\bibitem{caf98}
T. Cretegny, S. Aubry and S. Flach, Physica D {\bf 119}, 73 (1998).

\bibitem{fmf03}
S. Flach, A. E. Miroshnichenko and M. V. Fistul,
CHAOS {\bf 13}, 596 (2003).

\bibitem{ma94}
R. S. MacKay and S. Aubry, Nonlinearity {\bf 7}, 1623 (1994).

\bibitem{lsm97}
R. Livi, M. Spicci and R. S. MacKay,
Nonlinearity {\bf 10}, 1421 (1997).

\bibitem{ysk93}
Yu. S. Kivshar, Phys. Rev. E {\bf 48}, R43 (1993);
F. Fischer, Ann. Physik {\bf 2}, 296 (1993); Yu.A. Kosevich, Phys. Rev. B
{\bf 47}, 3138 (1993); Yu.A. Kosevich,
Phys. Rev. Lett. {\bf 71}, 2058 (1993).

\bibitem{sf95}
S. Flach, Phys. Rev. E {\bf 51}, 1503 (1995).

\bibitem{fdmf03}
S. Flach, J. Dorignac, A. E. Miroshnichenko and V. Fleurov,
Int. J. Mod. Phys. B {\bf 17}, 3996 (2003).

\bibitem{akk01}
S. Aubry, Ann. Inst. H. Poincare, Phys. Theor. {\bf 68}, 381 (1998);
S. Aubry, G. Kopidakis and V. Kadelburg,
Discrete and Continuous Dynamical Systems B {\bf 1}, 271 (2001).

\bibitem{gj01}
G. James, C. R. Acad. Sci. Paris {\bf 332}, 581 (2001).

\bibitem{sjca04}
B. Sanchez-Rey, G. James, J. Cuevas and J. F. R. Archilla,
Phys. Rev. B {\bf 70}, 014301 (2004).

\bibitem{fmff03}
S. Flach, A. E. Miroshnichenko. V. Fleurov and M. V. Fistul,
Phys. Rev. Lett. {\bf 90}, 084101 (2003).

\bibitem{uf61}
U. Fano, Phys. Rev. {\bf 124}, 1866 (1961);
J. A. Simpson and U. Fano, Phys. Rev. Lett. {\bf 11}, 158 (1963).

\bibitem{ma96}
J. L. Marin and S. Aubry, Nonlinearity {\bf 9}, 1501 (1996).

\bibitem{deft01}
B. Dey, M. Eleftheriou, S. Flach and G. P. Tsironis,
Phys. Rev. E {\bf 65}, 017601 (2001).

\bibitem{mk04}
M. Kastner, Phys. Rev. Lett. {\bf 92}, 104301 (2004).

\bibitem{fkm97}
S. Flach, K. Kladko and R. S. MacKay,
Phys. Rev. Lett. {\bf 78}, 1207 (1997).

\bibitem{movingfpu}
C. Claude, Yu. S. Kivshar, O. Kluth and K. H. Spatschek,
Phys. Rev. B {\bf 47}, 14228 (1993); F. Fischer,
Phys. Lett. A {\bf 182}, 417 (1993);
R. Khomeriki, S, Lepri and S. Ruffo, Physica D {\bf 168}, 152 (2002);
Yu. A. Kosevich and G. Corso, Physica D {\bf 170}, 1 (2002).

\bibitem{movingpt}
R. S. MacKay and J.-A. Sepulchre, J. Phys. A {\bf 35}, 3985 (2002).

\bibitem{fk99}
S. Flach and K. Kladko, Physica D {\bf 127}, 61 (1999);
J. Szeftel, G. X. Huang and V. Konotop, Physica D {\bf 181}, 215 (2003)

\bibitem{tp96}
S. Takeno and M. Peyrard, Physica D {\bf 92}, 140 (1996).

\bibitem{fkt97}
S. Flach, K. Kladko and S. Takeno, Phys. Rev. Lett.
{\bf 79}, 4838 (1997).

\bibitem{statistics}
Y. Zolotaryuk and J. C. Eilbeck,
J. Phys. Cond. Mat. {\bf 10}, 4553 (1998);
T. Cretegny, R. Livi and M. Spicci,
Physica D {\bf 119}, 88 (1998);
T. Cretegny, T. Dauxois, S. Ruffo and A. Torcini,
Physica D {\bf 121}, 109 (1998);
A. Sauerzapf and M. Wagner,
Physica B {\bf 263}, 723 (1999);
V. V. Mirnov, A. J. Lichtenberg and H. Guclu,
Physica D {\bf 157}, 251 (2001);
F. Piazza, S. Lepri and R. Livi,
J. Phys. A {\bf 34}, 9803 (2001);
R. Khomeriki, S. Lepri and R. Livi,
Phys. Rev. E {\bf 64}, 056606 (2001);
R. Khomeriki, Phys. Rev. E {\bf 65}, 026605 (2002);
R. Reigada, A. Sarmiento and K. Lindenberg,
Phys. Rev. E {\bf 66}, 046607 (2002);
R. Reigada, A. Sarmiento and K. Lindenberg,
Physica A {\bf 305}, 467 (2002);
L.S. Schulman, E. Mih\'okova, A. Scardicchio, P. Facchi,
M. Nikl, K. Pol\'ak, and B. Gaveau, Phys. Rev. Lett. {\bf 88},
224101 (2002);
H. B. Li, Phys. Lett. A {\bf 317}, 406 (2003);
Y. Doi, Phys. Rev. E {\bf 68}, 066608 (2003);
R. Reigada, A. Sarmiento and K. Lindenberg, 
CHAOS {\bf 13}, 646 (2003).














\end{thebibliography}
\end{document}